\documentclass[prl,aps,twocolumn,floatfix,amsmath,amssymb,superscriptaddress,tightenlines]{revtex4}
\usepackage{graphicx}% Include figure files
\usepackage{amsfonts}
\usepackage{bm}
\begin{document}

\date{\today}
\title{Scaling in the Fan of an Unconventional Quantum Critical Point}
\author{Roger G. Melko}
\affiliation{Department of Physics and Astronomy, University of Waterloo, Ontario, N2L 3G1, Canada} 
\affiliation{Materials Science and Technology Division, Oak Ridge National Laboratory, 
Oak Ridge TN, 37831}

\author{Ribhu K. Kaul}
\affiliation{Department of Physics, Harvard University, Cambridge MA, 02138}

\begin{abstract}
We present results of extensive finite-temperature Quantum Monte Carlo simulations on a SU(2) symmetric $S=1/2$ quantum antiferromagnet with a four-spin interaction~[Sandvik, Phys. Rev. Lett. {\bf 98}, 227202 (2007)]. 
Our simulations, which are free of the sign-problem and carried out on lattices containing in excess of $1.6 \times 10^4$ spins, indicate that the four-spin interaction destroys the N\'eel order at an unconventional $z=1$
 quantum critical point, producing a valence-bond solid paramagnet. Our results are  consistent with the `deconfined quantum criticality' scenario.
\end{abstract}
\maketitle

\newpage

Research into the possible ground states of SU(2) symmetric quantum antiferromagnets has thrived
over the last two decades, motivated to a large extent by the undoped parent compounds of the cuprate superconductors.  
In these materials, the Cu sites can be well described as $S=1/2$ spins on a two-dimensional (2D) square lattice that interact with an anti-ferromagnetic exchange, the archetypal model for which is the Heisenberg model. 
By now, it is well established~\cite{NeelRMP} that the ground state of this model 
with nearest-neighbor interaction has N\'eel order that spontaneously breaks the SU(2) 
symmetry. 
Two logical questions immediately arise: What possible paramagnetic ground states can be reached by tuning competing interactions that destroy the N\'eel state?  Are there universal quantum-critical points (QCP) that separate these paramagnets from the N\'eel phase? 

\begin{figure}
{
\includegraphics[width=2.7in]{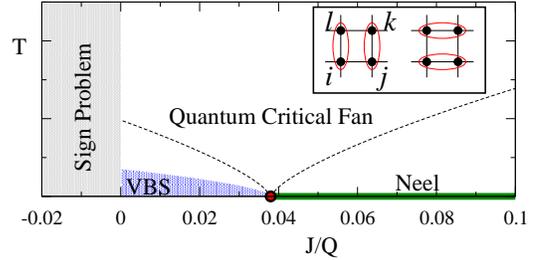}
\caption{(color online) 
Schematic of the proposed $T$-$J/Q$ phase diagram of the JQ model. 
Large-scale finite-T simulations presented here substantiate the following:
(i) The $T=0$ N\'eel order present for $J/Q \gg 1$ is destroyed at a QCP  ($J/Q\approx 0.038$); 
(ii) In the `quantum critical fan', there is scaling behavior characteristic of a $z=1$ QCP; 
(iii) An accurate estimate of the scaling dimension of the N\'eel field establishes that this transition is not in the $O(3)$ universality class; and,
(iv) The paramagnetic ground state for sufficiently small $J/Q$ is a VBS.
In the QMC basis used here, the region with $Q<0$ is sign problematic.
The inset shows how the frustrating $Q$ term is written in terms of bonds on a plaquette.
\label{fig:pd}}}
 \end{figure}

An answer to the first question is to disorder the N\'eel state by the proliferation of topological defects in the N\'eel order parameter~\cite{haldane88}. It was shown by Read and Sachdev~\cite{rs} that the condensation of these defects in the presence of quantum Berry phases 
results in a four-fold degenerate paramagnetic ground state, which breaks square-lattice symmetry due to the formation of a crystal of valence bonds
-- a valence-bond solid (VBS) phase. 
An answer to the second question was posed in recent work by 
Senthil {\it et al.}\cite{DQCP12}, where the possibility of a 
direct continuous N\'eel-to-VBS transition was proposed.
The natural field theoretic description of this `deconfined quantum critical point' is written in terms of 
certain fractionalized fields that are confined on either side of the QCP and become `deconfined' precisely at the critical point. As is familiar from the general study of QCPs, these fractional excitations are expected to influence the physics in a large fan-shaped region that extends above the critical point at finite-$T$~\cite{subirbook} (see Fig.~\ref{fig:pd}).

It is clearly of great interest to find models that harbor a direct N\'eel-VBS QCP and that can be studied without approximation on large lattices.  Currently, the best candidate is the `JQ' model, introduced by Sandvik \cite{JQ_1}, which
is an $S=1/2$, SU(2) invariant antiferromagnet with a frustrating four-spin interaction,
\begin{equation}
\label{eq:jq}
H_{\rm JQ}=J\sum_{\langle ij \rangle} {\bf S}_i \cdot {\bf S}_j
-Q\sum_{\langle ijkl \rangle} ({\bf S}_i \cdot {\bf S}_j-\frac{1}{4})
({\bf S}_k \cdot {\bf S}_l-\frac{1}{4}),
\end{equation}
where indices are arranged as in the inset of Fig.~\ref{fig:pd}.
Using a $T=0$ projector Quantum Monte Carlo (QMC) method on lattices sizes up to $32\times 32$ \cite{JQ_1},
Sandvik showed that the four-spin interaction destroys N\'eel order and produces a VBS phase at $J/Q \sim 0.04$.
Close to this critical value of $J/Q$, scaling in the spin and dimer
correlation functions suggests a continuous transition with anomalous dimensions of the N\'eel and 
 VBS order parameters equal, with a common value $\eta = 0.26(3)$.
In this Letter, we explore the candidate N\'eel-VBS QCP in the full $T-J/Q$ phase diagram on large lattices using a complementary finite-$T$ QMC technique, the Stochastic Series Expansion (SSE) method with directed loops~\cite{DIRloop}.  The SSE QMC allows access 
to the physically important quantum critical fan (see Fig.~\ref{fig:pd}),
and admits high-accuracy estimates for the spin stiffness, $\rho_S$, and the uniform susceptibility, $\chi_u$.  The scaling of these observables provides
strong evidence for a continuous $z=1$ transition in the JQ model.

{\it Basis and Sign of Matrix Elements:}  A priori, it is unclear that SSE simulations of $H_{\rm JQ}$ are free of the notorious sign-problem: a fluctuating sign in the weights used in the QMC sampling. 
In the SSE, 
finding an orthogonal basis in which all 
off-diagonal matrix elements of the Hamiltonian are non-positive solves the sign-problem. 
%In the $S^z$ basis, this would seem to preclude even the nearest-neighbor Heisenberg  antiferromagnet ($J>0$ and $Q=0$) from being sign-problem-free. 
A simple unitary transformation on the $S^z$ basis (a $\pi$-rotation about the $z$-axis on one sub-lattice) results in a new basis in which,  for $J,Q>0$, all off-diagonal matrix elements of $H_{\rm JQ}$ are non-positive, allowing sign-problem free simulations (Fig.~\ref{fig:pd}). We note that this non-positivity condition is also the main ingredient in the proof of the {\em Marshall sign theorem}, allowing us to infer that the ground state of $H_{\rm JQ}$ for $J,Q>0$ must be a spin-singlet. As shown below, this singlet state changes from N\'eel at $Q\ll J$ to VBS at $J\ll Q$.

\begin{figure}[]
{
\includegraphics[width=3.2in]{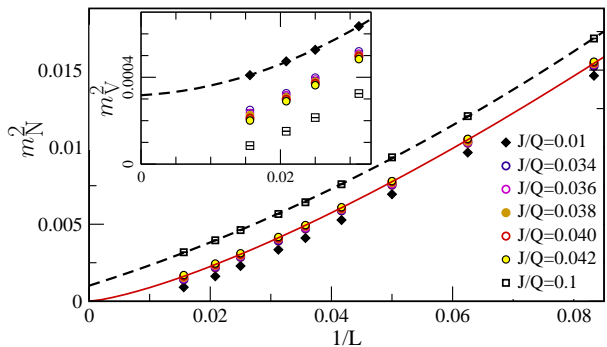}
\caption{(color online) $T \rightarrow 0$ converged N\'eel (main) and VBS (inset) order parameters as a function of $1/L$.  
%The N\'eel condensate (main) is suppressed as the critical coupling is approached from above. 
Dashed lines are
quadratic fits that illustrate the finite condensate in the ordered phases.  The solid (red) line is a fit to the form $y=c_1 x^{c_2}$ (illustrated for $J/Q=0.040$), where 
$c_2=z+\eta_{\rm N}$ is expected at the critical coupling.  In fitting to the nine $L$ values for each $J/Q$, we find a minimum in the chi-squared value (per degree of freedom) of 3.1 for $J/Q=0.040$, with $c_2\approx 1.35(1)$.  For $J/Q=0.038$, the chi-squared value is 3.9, with $c_2\approx 1.37(1)$. All other $J/Q$ produce much larger chi-square (greater than 10).  
%The inset shows the VBS order parameter, which is suppressed as the critical coupling is approached from below.
\label{T0FSS}} }
\end{figure}

\begin{figure}[]
{
\includegraphics[width=3.2in]{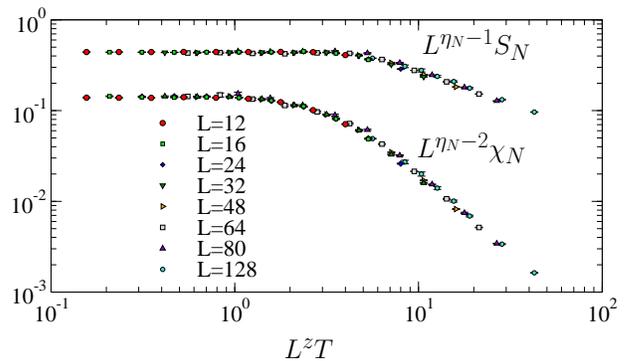}
%\caption{(color online) Criticality of the N\'eel  field at $J=0.038$: collapse of the N\'eel structure factor ($S_N$) and susceptibility ($\chi_N$) with $z=1$ and $\eta_N=0.35$, determining the universal functions $\mathbb{X}_{\rm S}(x)$ and $\mathbb{X}_{\rm \chi}(x)$ (up to non-universal scale factors on the $x$ and $y$ axes). $\eta_N$ (the anomalous dimension of the N\'eel field) is the only fit parameter for the collapse of $S_N$ and $\chi_N$ over roughly three orders of magnitude.
\caption{(color online) Criticality of the N\'eel  field at $J=0.038$: collapse of the N\'eel structure factor ($S_N$) and susceptibility ($\chi_N$) with $z=1$ and $\eta_N=0.35$, determining the universal functions $\mathbb{X}_{\rm S}(x)$ and $\mathbb{X}_{\rm \chi}(x)$ (up to non-universal scale factors on the $x$ and $y$ axes). The only fit parameter for both $S_N$ and $\chi_N$ is $\eta_N$, the anomalous dimension of the N\'eel field. 
\label{fig:etaN}} } 
\end{figure}

{\it Numerical Results:} Using the SSE QMC, we studied various physical observables in the JQ model
on finite-size lattices of linear dimension $L$ (with number of spins $N_{\rm spin}=L^2$).
Particular attention was paid to the scaling of the  spin stiffness $\rho_s={\partial^2 E_0}/{\partial \phi^2}$ ($E_0$ is the energy and $\phi$ is a twist in the boundary conditions) and
the uniform spin susceptibility $\chi_u=\langle (\sum_i S^z_i)^2 \rangle/T N_{\rm spin}$. In the $S^z$ basis used here, it is easy to measure the correlation functions $C^z_{\rm N}({\bf r},\tau)= \langle S^z({\bf r},\tau) S^z(0,0) \rangle$ and $
C^z_{\rm V}({\bf r},\tau)= \langle [ S^z({\bf r},\tau) S^z({\bf r}+\hat{{\bf x}},{\bf \tau}) ] [S^z(0,0) S^z(\hat{{\bf x}},0)] \rangle$. While $C^z_{\rm N}$ is the correlation function of the N\'eel order parameter, the VBS  order is indicated by $C^z_{\rm V}$, which is the correlation function of the composite operator $S^z({\bf r})S^z({\bf r+\hat{ x}})$,
receiving contribution from both the 
%usual singlet-VBS order parameter ${\bf S}({\bf r}) \cdot {\bf S}({\bf r+\hat{ x}})$,
standard VBS order parameter ${\bf S}({\bf r}) \cdot {\bf S}({\bf r+\hat{ x}})$
%\cite{JQ_1},
as well as the traceless symmetric tensor constructed from $S^i({\bf r})S^j({\bf r+\hat{ x}})$. 
%The mixing-in of order parameters in $C^z_{\rm V}$ turns out to make it difficult for us to extract the criticality of the VBS order parameter accurately; however, as shown below, $C^z_{\rm V}$ is still useful to detect the VBS ordering.   
Structure factors for the N\'eel and VBS phases are constructed from these correlation functions by Fourier transformation at equal time, $S_{\rm N,V}[{\bf q}]=\sum_{\bf r} [\exp(- i {\bf q} \cdot {\bf r}) C^z_{\rm N,V}({\bf r,\tau=0}) ]/N_{\rm spin}$, from which the order parameters are defined at the observed
ordering wavevectors:
$m^2_{\rm N,V}=S_{\rm N,V}[{\bf q}_{\rm N,V}]/N_{\rm spin}$.
Zero-frequency susceptibilities  ($\chi_{\rm N}$ and $\chi_{\rm V}$) are constructed by integrating over all $\tau$ and Fourier transforming in space to the ordering vectors, ${\bf q}_{\rm N,V}$. 

Examination of the full ${\bf q}$-dependent structure factors indicate the presence of sharp ordering
wavevectors in $S_{\rm N}[{\bf q}_{\rm N}=(\pi,\pi)]$ for large $J/Q$ and $S_{\rm V}[{\bf q}_{\rm V}=(\pi,0)$ or $(0,\pi)]$ 
(the latter in the case where the correlator is measured with ${\hat{\bf y}}$) for large $Q/J$ \cite{VBSs}, 
confirming the N\'eel and VBS phases observed in Ref.~\cite{JQ_1}.  
As shown in Fig.~\ref{T0FSS}, $T\rightarrow0$ converged data scales convincingly to 
a non-zero value for $m^2_{\rm N} $ at $J/Q=0.1$ and for $m^2_{\rm V} $ at $J/Q=0.01$. 
The critical coupling appears to occur between $J_c\approx0.038$ and 0.040 (we set $Q=1$ fixed throughout), such that as $J_c$ is approached from above (below) the extrapolated N\'eel 
(VBS) order parameter is suppressed.  Very near $J_c$,  both order parameters vanish within our error bars, while a power law with no $y$-intercept
fits the N\'eel data with high accuracy. 
More specifically, at $J_c$, scaling arguments require $S_N\propto L^{1-\eta_N}\mathbb{X}_{\rm S}(L^zT/c)$ and $\chi_N\propto L^{2-\eta_N}\mathbb{X}_\chi (L^zT/c)$, with $\eta_N$ the anomalous dimension of the N\'eel field. In Fig.~\ref{fig:etaN}, we verify this scaling behavior and determine the universal functions $\mathbb{X}_{\rm S}$ and $\mathbb{X}_\chi$. 
Both analyses illustrated in Figs.~(\ref{T0FSS},\ref{fig:etaN}) give a consistent estimate of $\eta_N\approx 0.35(3)$. This value is larger than the result of $\eta_N\approx 0.26(3)$ from Ref.~\cite{JQ_1}. While the exact source of this discrepancy is unclear due to the entirely different methods used to extract the exponents, we note that (a) our analysis does not involve extra fit parameters from the inclusion of sub-leading corrections, and (b) the collapse of both $S_N$ and $\chi_N$ takes place over two and a half orders of magnitude of $LT$ with only one common fit parameter, $\eta_N$; both facts give us confidence in our estimate. 
The critical scaling of $C^z_{\rm V}$ is more complicated; due to the
aforementioned mixing-in of two order parameters, $C^z_{\rm V}$ is expected to receive two {\em independent} power-law contributions.  Indeed, it is difficult to disentangle these individual contributions on the limited range of lattices sizes available, precluding us from verifying the 
 proposal~\cite{JQ_1}  that $\eta_{\rm N} = \eta_{\rm V}$.

We now turn to 
%our most convincing results for a continuous transition -- 
an 
analysis of the scaling properties of $\chi_u$ and $\rho_s$ in the hypothesized quantum critical fan region of Fig~\ref{fig:pd}.  $\chi_u$ and $\rho_s$, being susceptibilities of conserved quantities have no anomalous scaling dimension, and hence at finite-$T$ and $L$ in the proximity of a scale-invariant critical point, assuming hyper-scaling:
\begin{eqnarray}
\label{eq:rho}
\rho_s (T,L, J) &=& \frac{T}{L^{d-2}} \mathbb{ Y} \left( \frac{L^z T}{c}, g L^{1/\nu} \right),\\
\label{eq:chi}
\chi_u (T,L,J) &=& \frac{1}{T L^d} \mathbb{Z} \left(\frac{L^z T}{c}, g L^{1/\nu}\right),
\end{eqnarray}
where $g \propto (J-J_c)/J_c$. At criticality ($g=0$), it is easy to see that $\mathbb{Y}(x\rightarrow 0,0)=\mathcal{A}_\rho/x$ and ${\mathbb{Z}}(x\rightarrow \infty,0)=\mathcal{A}_\chi x^{d/z}$, where $\mathbb{Y}(x,y)$ and $\mathbb{Z}(x,y)$ are universal scaling functions and $\mathcal{A}_\chi, \mathcal{A}_\rho$ are universal amplitudes of the quantum critical point; $c$ is a non-universal velocity. 

\begin{figure}[]
{
\includegraphics[width=3.2in]{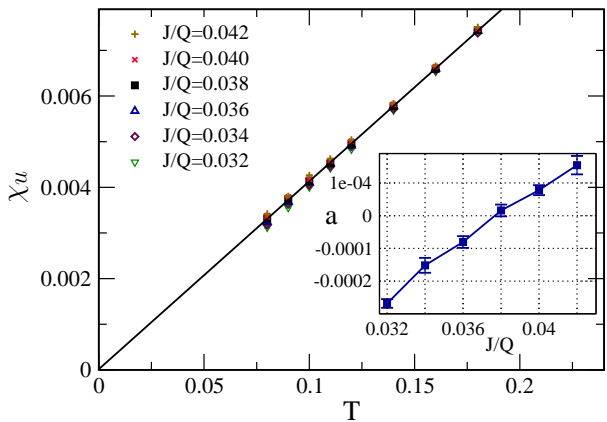}
\caption{(color online) 
Finite-$T$ uniform susceptibility, for a $L=128$ system near $J_c$.
Error bars are much smaller than the symbol size.  %, typically around $1\times 10^{-5}$.
For the region $0.08 \leq T \leq 0.18$, the data is highly linear, and a straight-line
fit for $J/Q=0.038$ (shown) intercepts the origin within error bars.  
Intercepts of straight-line fits for all data sets are in the inset. From the slope of the linear-$T$ behavior we obtain $\mathcal{A}_\chi/c^2 = 0.0412(2)$.
\label{chiT}} }
\end{figure}
\begin{figure}
{\includegraphics[width=3.6in]{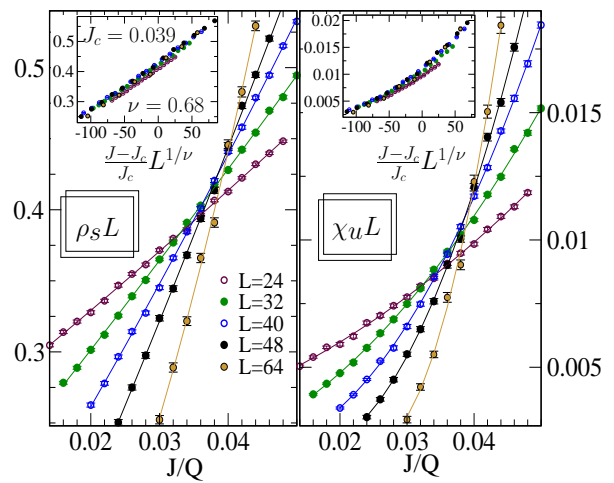}
\caption{(color online)
Zoom-in of the spin stiffness and susceptibility close to the expected critical point, taken for simulations cells of size $L=1/T$.  
Data in the inset is scaled to 
get the best collapse for the largest system sizes, which occurs for 
$J_c \sim 0.039(1)$ and $\nu \sim 0.68(4)$.
\label{RhoChi_C}} }
\end{figure}
At criticality and $L\rightarrow \infty$, one can show from Eq.~(\ref{eq:chi}) that $\chi_u = \frac{\mathcal{A}_\chi}{ c ^{d/z}}T^{d/z -1}$; i.e.~for a $z=1$ transition, $\chi_u$ should be $T$-linear and have a zero intercept on the y-axis at $T=0$~\cite{Chubukov}.
In Fig.~\ref{chiT}, $\chi_u$ data for an $L=128$ system is presented. Within our error bars, this data is $L\rightarrow \infty$ converged for the region of $T$ shown; at smaller $T$ the finite-size gap causes an exponential reduction in $\chi_u$. %Close to $J=0.038$ the data fits the form $a+bT$,
 The inset shows how the extracted value of the y-intercept, $a$ (from a fit to the form $a+bT$), changes sign as the coupling is tuned, 
consistent with $0.036 \leq J_c \leq 0.040$ and demonstrating to high precision the $z=1$ scaling. 

Turning to study Eqs.~(\ref{eq:rho},\ref{eq:chi}) further, one may hold the first argument of the universal functions fixed by setting $L=1/T$ (assuming $z=1$ as indicated above).  
In order to achieve this, we performed extensive simulations on lattices sizes up to $L=1/T=64$, illustrated
in Fig.~\ref{RhoChi_C}.  According to Eqs.~(\ref{eq:rho},\ref{eq:chi}),  data curves for $L \rho_s$ and $L \chi_u$  plotted versus $J$ should show a crossing point with different $L$ precisely at $J_c$.  We find that for  
relatively large sizes ($32 \leq L \leq 64$) the crossing point converges quickly in the interval 
$0.038 \leq J \leq 0.040$. The insets show the data collapse when the $x$-axis is re-scaled to $gL^{1/\nu}$ (with $\nu=0.68$).  We note that with the inclusion of small sub-leading corrections
(of the form $a_\omega/L^\omega$), the crossing point and data collapse of $\rho_s$ and $\chi_u$ can be 
made consistent, at the expense of two more fit parameters, even for much smaller system sizes than illustrated~\cite{JQ_todo}. 
In contrast to the U(1)
symmetric JK model \cite{JKexp_1}, where the absence of a $T$-linear $\chi_u$ and a crossing in the data for $\rho_s L$
cast doubt on its interpretation as a $z=1$ QCP, the present data for this SU(2) symmetric model
gives strong support for a $z=1$ QCP  between $0.038 \leq J \leq 0.040$.

\begin{figure}
{\includegraphics[width=3.1in]{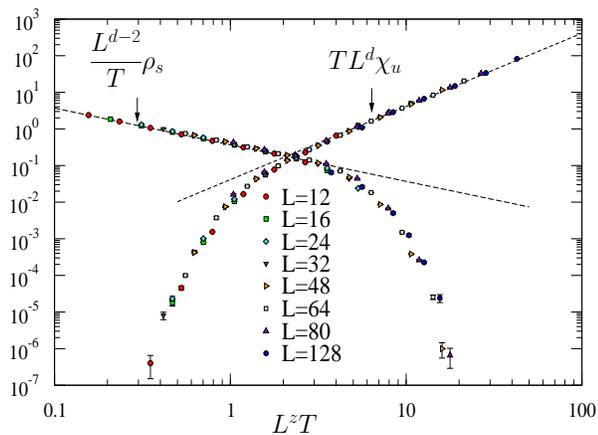}
\caption
{
Scaling of $\chi_u$ and $\rho_s$ at $J=0.038\approx J_c$, with $z=1$ and $d=2$. These plots are the universal functions $\mathbb{Y}(x,0)$ and $\mathbb{Z}(x,0)$ up to the non-universal scale factor $c$ on the $x$-axis. 
The expected asymptotes (see text) are plotted as dashed lines  $\mathbb{Y}(x\rightarrow 0,0)=\mathcal{A}_\rho/x$ and ${\mathbb{Z}}(x\rightarrow \infty,0)=\mathcal{A}_\chi x^{d/z}$. From fits to the data, we find $\mathcal{A}_\chi/c^2=0.041(4)$ and $\mathcal{A}_\rho c=0.37(3)$, allowing us to estimate a universal model-independent number associated with the QCP, $ \mathcal{A}_\rho \sqrt{\mathcal{A}_\chi}\approx 0.075(4)$.
\label{collapse}
} 
}
\end{figure}

Finally, we hold the second argument of the scaling functions [Eqs.~(\ref{eq:rho},\ref{eq:chi})] constant by tuning the system to $g=0$.  One then expects a data collapse for $\rho_s/T$ and $L\chi_u$ when they are plotted as a function of $L^z T$ (with $z=1$). Fig.~\ref{collapse} shows this collapse for simulations 
carried out with extremely anisotropic arguments $LT$, varying over almost three orders of magnitude. There is an excellent data collapse over {\em $8$ orders of magnitude} of the range of the universal functions, with no fit parameters. 
This data together with that in Fig.~\ref{chiT} provide our most striking evidence for the existence of a QCP with $z=1$ in the proximity of $J/Q \approx 0.038$.

{\em Discussion:} In this paper we have presented extensive data for the SU(2) symmetric JQ model
which indicates that the N\'eel order (present when $J\gg Q$) is destroyed at a continuous quantum transition as $Q$ is increased \cite{JQ_1}.
In the finite-$T$ quantum critical fan above this QCP, scaling behavior is found that confirms
the dynamic scaling exponent $z=1$ to high accuracy. 
The anomalous dimension of the N\'eel field at this transition is determined to be 
$\eta_{\rm N} \approx 0.35(3)$, almost an order of magnitude more than its value of $0.038$ \cite{Camp} for a conventional $O(3)$ transition. For sufficiently large values of $Q$ we find that the system enters a spin-gapped phase with VBS order. 
To the accuracy of our simulations, our results are fully consistent with a direct continuous
QCP between the N\'eel and VBS phases, with a critical coupling between $J/Q\approx0.038$ 
and $J/Q\approx0.040$.  Although our finite size study cannot categorically rule out a weak first-order transition, we have found no evidence for double-peaked distributions, indicating an absence of this sort of first-order behavior on the relatively large length scales studied here.
It is interesting to compare our results to 
the only theory currently available for a continuous N\'eel-VBS transition: the deconfined quantum criticality scenario~\cite{DQCP12}, in which the N\'eel-VBS transition is described by the non-compact $\mathbb{CP}^1$ field theory.
All of the qualitative observations above, including an unusually large $\eta_{\rm N}$~\cite{motrunich} agree with the predictions of this theory. Indeed, our estimate of $\eta_{\rm N} \approx 0.35$ [Fig.~\ref{fig:etaN}] is in remarkable numerical agreement with a recent field-theoretic computation~\cite{nazsant} of this quantity, which finds $\eta_{\rm N}=0.3381$.
With regard to other detailed quantitative comparisons, we have provided the first step by computing many universal quantities, $\mathbb{X}_\chi(x)$, $\mathbb{X}_S(x)$, $\mathbb{Y}(x,0)$,  $\mathbb{Z}(x,0)$ and $ \mathcal{A}_\rho \sqrt{\mathcal{A}_\chi}\approx 0.075$  [Fig.~\ref{collapse}] in the JQ model. Analogous computations in the $\mathbb{CP}^1$ model, although currently unavailable~\cite{kuklov} are highly desirable to further demonstrate that the JQ model realizes this new and exotic class of quantum criticality.

We acknowledge scintillating discussions with S.~Chandrasekharan, A.~del Maestro, T.~Senthil, and especially S.~Sachdev and A.~Sandvik.
This research (RGM) was sponsored by D.O.E. contract
DE-AC05-00OR22725.
%with ORNL, managed by UT-Battelle, LLC, 
RKK acknowledges financial support from NSF DMR-0132874, DMR-0541988 and DMR-0537077. Computing resources 
were contributed by NERSC (D.O.E. contract DE-AC02-05CH11231), NCCS, the HYDRA cluster at Waterloo, and the DEAS and NNIN clusters at Harvard.

\bibliographystyle{apsrev}
%\bibliography{rmBiblio}

\begin{thebibliography}{99}

\bibitem{NeelRMP} E. Manousakis, Rev. Mod. Phys. {\bf 63}, 1 (1991). 

\bibitem{haldane88}F.~D.~M.~Haldane, Phys. Rev. Lett. {\bf 61}, 1029 (1988)
\bibitem{rs} N. Read and S. Sachdev, Phys. Rev. B {\bf 42}, 4568 (1990).

\bibitem{DQCP12} T. Senthil {\em et al.}
%, A. Vishwanath, L. Balents, S. Sachdev, and M. P. A. Fisher, 
Science {\bf 303}, 1490 (2004); 
%T. Senthil, L. Balents, S. Sachdev, A. Vishwanath, and M. P. A. Fisher, 
Phys. Rev. B {\bf 70}, 144407 (2004).

\bibitem{subirbook} S. Sachdev, {\it Quantum Phase Transitions} (Cambridge University Press, New York, 1999). 

\bibitem{JQ_1} A. W. Sandvik, Phys. Rev. Lett. {\bf 98}, 227202 (2007).

\bibitem{DIRloop} O.~F.~Sylju{\aa}sen and A.~W.~Sandvik, Phys. Rev. E {\bf 66}, 046701 (2002);
R.~G.~Melko and A.~W.~Sandvik, Phys. Rev. E {\bf 72}, 026702 (2005).

\bibitem{VBSs}
The VBS order is also visible in measurements of correlation functions between off-diagonal terms~\cite{JQ_todo}.

\bibitem{JQ_todo} R. K. Kaul and R. G. Melko, {\it unpublished}.


\bibitem{Chubukov}  A. V. Chubukov {\em et al.}, 
%S. Sachdev, and J. Ye, 
Phys. Rev. B  {\bf 49}, 11919 (1994).

\bibitem{JKexp_1} A. W. Sandvik and R. G. Melko, cond-mat/0604451 (2006);
Ann. Phys. (NY), {\bf 321}, 1651 (2006).

\bibitem{Camp} M. Campostrini {\em et al.}, %M. Hasenbusch, A. Pelissetto, P. Rossi, and E. Vicari, 
Phys. Rev. B {\bf 65}, 144520 (2002).

\bibitem{motrunich}
O. I. Motrunich and A. Vishwanath, Phys. Rev. B {\bf 70}, 075104 (2004).

\bibitem{nazsant}
Z. Nazario and D. I. Santiago, %Phys. Rev. Lett. {\bf 97}, 197201 (2006); 
Nucl. Phy. B {\bf 761}, 109 (2007)


\bibitem{kuklov} A. Kuklov {\em et al.}, 
%N. Prokof'ev, B. Svistonuv, and M. Troyer, 
Annals of Physics {\bf 321}, 1602 (2006); 
found a discontinuous transition in a U(1) deformation of the $\mathbb{CP}^1$ model. Results relevant to the SU(2) invariant $H_{\rm JQ}$ are so far unavailable (see however~\cite{motrunich}).  


\end{thebibliography}

\end{document}